\documentclass{elsart}

\usepackage{epsfig}

\usepackage{amssymb}

\begin{document}

\begin{frontmatter}

\title{On anomalous distributions in intra-day financial time series and
Non-extensive Statistical Mechanics}

\author{Silvio M. Duarte Queir\'{o}s}

\ead{sdqueiro@cbpf.br}
\address{
Centro Brasileiro de Pesquisas F\'\i sicas, \\
Rua Dr.  Xavier Sigaud 150,22290-180 Rio de Janeiro-RJ, Brazil
}

\begin{abstract}
In this paper one studies the distribution of log-returns (tick-by-tick)
in the Lisbon stock market and shows that it is well adjusted by the
solution of the equation, \mbox{$\frac{dp_{x}}{d\left| x\right| }=-\beta
_{q^{\prime }}p_{x}^{q^{\prime }}-\left( \beta _{q}-\beta _{q^{\prime
}}\right) p_{x}^{q}$}, which corresponds to a generalization of the
differential equation which has as solution the power-laws that optimise the
entropic form $S_{q}=-k \frac{1-\int p_{x}^{q}\ dx}{1-q}$, base of present
non-extensive statistical mechanics.
\end{abstract}

\end{frontmatter}

\section{Introduction}
Since Mandelbrot's work \cite{mandelbrot} on analysing cotton prices that L\'{e}vy distribution has been
widely used, but, frequently with L\'{e}vy scaling coefficients greater than 
$2$, and so, out of L\'{e}vy regime\cite{stanley,pobodnik}. An alternative
way of describing these tailed distributions can be power laws (usually
called ${\it q}${\it -Gaussians }or {\it Tsallis distributions}) obtained by
optimising the entropic form proposed by Constantino Tsallis\cite
{silvio,ausloos,tsallis-borland}. Those distributions, beyond presenting the
advantage of finite second (and higher) order momenta (for a certain range
of values of its characterizing index $q$) are also the solution of various
kinds of non-linear Fokker-Planck equations\cite{abe,borland-pre,richmond}
often used as continuous time approaches to financial systems. In this
manuscript one shows that a generalisation of $\ $non-extensive formalism
permits a good adjustment to the tick-by-tick log-return distributions for
Lisbon Stock Market.
\section{From Boltzmann-Gibbs-Shannon to anomalous distribution}
It is known that the consecrated Boltzmann-Gibbs-Shannon (BGS) entropic
form, $S=-\int p\left( x\right) \ \log p\left( x\right) \ dx$, is able to
provide a perfect description of an all range of phenomena with microscopic 
spatial/temporal short-range interactions, but is not capable
to give proper answers when systems present microscopic spatial/temporal 
long-range interactions \cite{tamarit,tsallis-celia} or
multifractal structure \cite{baldovin}. In order to deal with this glitch,
Tsallis presented the entropic form\cite{tsallis-sq}, 
\begin{equation}
S_{q}=-k \frac{1-\int \left[ p\left( x\right) \right] ^{q}\ dx}{1-q}
\end{equation}
$\left( q\;\epsilon \;\Re \right) $, which contains BGS as a
particular case when $q=1$. For this entropic form, the optimisation,
under the appropriate constrains $\int p\left( x\right) \ dx=1$ (unitary norm) 
and $\int
x^{2}\frac{\ \left[ p\left( x\right) \right] ^{q}}{\int \left[ p\left(
x\right) \right] ^{q}dx}\ dx=\sigma _{q}^{2}$ (finite generalised variance), 
yields the following power-law ($q$-Gaussian), 
\begin{equation}
p\left( x\right) \;\sim \;\left[ 1-\beta _{q}\left( 1-q\right)
x^{2}\right] ^{\frac{1}{1-q}}\equiv e_{q}^{-\beta _{q}x^{2}},
\label{q-gauss1}
\end{equation}
which in the limit $q\rightarrow 1$ recovers the Gaussian distribution. The
$q$-Gaussian, $e_{q}^{-\beta _{q}x^{2}}$, can be rewritten, in a quite
interesting way, as the solution for the non-linear differential equation,
\begin{equation}
\frac{dp\left( x\right) }{d\left( x^{2}\right) }=-\beta _{q}\left[ p\left(
x\right) \right] ^{q}.  \label{q-gauss2}
\end{equation}
From this equation one can obtain the (simmetrical)$\;q$-exponential
distribution changing $x^{2}$ by $\left| x\right| $ and the exponential
distribution considering also $q\rightarrow 1$.
For reasons that will be presented in the next section let one extends Eq. (%
\ref{q-gauss2}), for the $q$-exponential case, in the same way that was done in
previous works on protein re-association \cite{bemski} and flux of cosmic rays, 
by introducing another index $q^{\prime }$, 
\begin{equation}
\frac{dp\left( x\right) }{d\left| x\right| }=-\beta _{q^{\prime }}\left[
p\left( x\right) \right] ^{q^{\prime }}-\left( \beta _{q}-\beta _{q^{\prime
}}\right) \left[ p\left( x\right) \right] ^{q}.  \label{gen1}
\end{equation}
(It is interesting to notice that if we consider $q^{\prime }=1$ and $q=2$
one obtains the expression used by Max Planck in 1900 that lead him to the
black-body radiation law \cite{max}). The solution of Eq. (\ref{gen1})
yields,
\begin{eqnarray}
\left| x\right|  &=&\frac{1}{\beta _{q^{\prime }}}\left\{ \frac{\left[
p\left( x\right) \right] ^{-\left( q^{\prime }-1\right) }}{q^{\prime }-1}-%
\frac{\left( \beta _{q}/\beta _{q^{\prime }}\right) -1}{1+q-2q^{\prime }}%
\times \right.  \\
&&\left. \times \left[ J\left( 1;q-2q^{\prime },q-q^{\prime },\left( \beta
_{q}/\beta _{q^{\prime }}\right) -1\right) -J\left( p\left( x\right)
;q-2q^{\prime },q-q^{\prime },\left( \beta _{q}/\beta _{q^{\prime }}\right)
-1\right) \right] \right\}   \nonumber
\end{eqnarray}
where $J\left( p\left( x\right) ;a,b,c\right) =\left[ p\left( x\right) %
\right] ^{1+a}F\left( \frac{1+a}{b};\frac{1+a+b}{c};\left[ p\left( x\right) %
\right] ^{b}c\right) $ and $F$ the hypergeometric function. For $1<q^{\prime }<q$
and $\beta _{q^{\prime }}\ll \beta _{q}$ three regions can be observed. The
first one (where $q$ governs) for $0\leq \left| x\right| \ll \left| x\right| ^{\ast }$, where $%
\left| x\right| ^{\ast }=\frac{1}{\left( q-1\right) \beta _{q}}$, a second
one (where both $q$ and $q^{\prime}$ govern) for $\left| x\right| ^{\ast }\ll \left| x\right| \ll \left| x\right|
^{\ast \ast }\equiv \frac{\left[ \left( q-1\right) \beta _{q}\right] ^{\frac{%
q^{\prime }-1}{q-q^{\prime }}}}{\left[ \left( q^{\prime }-1\right) \beta
_{q^{\prime }}\right] ^{\frac{q-1}{q-q^{\prime }}}}$ and the third one (governed by  $q^{\prime}$)  for $%
\left| x\right| \gg \left| x\right| ^{\ast \ast }$.
\section{Application to Lisbon Stock Market}
\begin{figure}[htb]
 \begin{center}
 \epsfig{figure=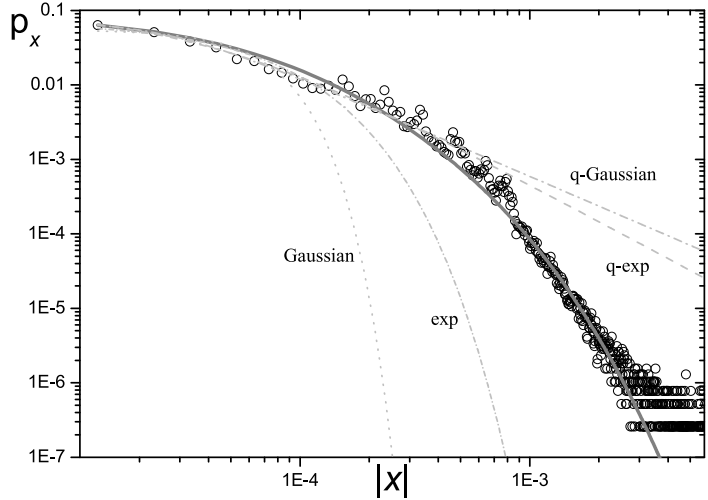,width=0.7\textwidth,clip=}
 \caption{{\protect\small Probability density function (PDF) for the tick by
tick log-return in the Lisbon Stock Market between 1}$^{st}${\protect\small %
\ Feb. 1996 to 28}$^{th}${\protect\small \ Jun. 2002 ( 4 million ticks
approx.). The full line represents the solution of Eq. (\ref{gen1}) and it is
clearly best approach to PDF from data (circles) when compared with the best Gaussian, }$q${\protect\small -Gausssian (}$q=2.51$%
{\protect\small ), exponential and q-exponential (}$q=1.59${\protect\small )
fits that are also plotted.}}
 \label{fig2}
 \end{center}
\end{figure}

Let one now considers the variable $x_{i}$ as the tick-by-tick log-return
obtained by, $x_{i}=\log \frac{\Pi _{i}}{\Pi _{i-1}}$,
with $\Pi _{i}$ representing the value of stock market index at tick $i$. Here
one neglets the time spacing between transactions which are, in mean, of
order of $10$ seconds and its possible influence in the magnitude of variations.
The description of log-return's distribution has been successfully done with $%
q$-Gaussians when the lag between elements at least of the order of minutes
(or greater), but fails when one deals with time series where lag is smaller
for which data seems to be better approached by exponential functions \cite
{yakovenko}. However, as can be seen in Fig. \ref{fig2}, for the case of the
tick-by-tick return in the Lisbon stock market the exponential function it
is only able to fit the central part of the distribution. By solving
numerically the Eq.(\ref{gen1}) with $q^{\prime }=1.076$, $q=1.534$, $\beta
_{q^{\prime }}=6.59\times 10^{3}$ and $\beta _{q}=7.47\times 10^{4}$ one
obtained the full line which presents a very good agreement, and clearly the
best compared to other distributions, for data from Lisbon stock market for
six decades in ordinate and two and a half decades in abcissa.
For the values $q,q^{\prime },\beta _{q}$ and $\beta _{q^{\prime }}$ used, $%
\left| x\right| ^{\ast }\simeq 0.00003$ and $\left| x\right| ^{\ast \ast
}\simeq 0.004$. Unfortunately, the finiteness of data does not allow the
observation of the second crossover which occurs at a value of $x$ where it
is not possible to have a good statistics.
\section{Concluding remarks}
In conclusion one can state that like in biological and cosmological
phenomena, the superposition of power-laws which maximise Tsallis'
entropy can be an excellent approach in order to describe anomalous
distributions in finance, fact that could be a sign of a peculiar dynamics
(probably due to market characteristics like volume, number of traders)
different than the ones presented until now. Despite the fact that
aggregations were not analised here, existing works, where two regions can
also be observed in the probability density function of returns \cite
{pobodnik}, points that this distribution will relax firstly to the $q$%
-Gaussian \cite{tsallis-borland} and finally to the Gaussian distribution 
\cite{silvio}. Although the study of tick-by-tick time series is usually
negleted in favour of, e.g.,1 minute lag time series, its analysis could be
important in the reach of a better understanding of the nature of these
(beautiful financial) systems, because all phenomena observed at major time
scales have their origin in this time scale dynamics.

\bigskip
The author is thankful to Constantino Tsallis and E.P. Borges for useful comments, Data
Services of Euronext Lisbon for free data providing and Funda\c{c}\~{a}o
para a Ci\^{e}ncia e Tecnologia (Portuguese agency) for financial support.
Financial support from APFA4 organising committee, which made his
participation possible, is deeply acknowledge.
\end{document}